\documentclass[10pt,aps,prl,twocolumn,superscriptaddress,floatfix]{revtex4-2}

\usepackage[utf8]{inputenc}
\usepackage{amsmath,amssymb}
\usepackage{graphicx}
\usepackage{bm}
\usepackage{color}
\usepackage{tikz}
\usepackage{hyperref}
\usepackage{float}

\begin{document}

\title{Area and Perimeter Rules of Velocity Circulation in Two-Dimensional Turbulence with Large-scale Absolute Equilibrium}

\author{Zi-Ang Zhang}
\affiliation{School of Mechanics and Engineering Science and State Key Laboratory for Turbulence and Complex Systems, Peking University, Beijing 100871, China}

\author{Jin-Han Xie}
\email{jinhanxie@pku.edu.cn}
\affiliation{School of Mechanics and Engineering Science and State Key Laboratory for Turbulence and Complex Systems, Peking University, Beijing 100871, China}

\date{\today}

\begin{abstract}

We demonstrate that the area rule of velocity circulation -- traditionally associated with the turbulent inertial range but shown not to be exact -- is strictly satisfied in the large-scale absolute equilibrium of two-dimensional (2D) homogeneous isotropic turbulence under enstrophy equipartition. We also derive a novel perimeter rule from the 2D inviscid loop equation, which posits that the probability distribution function (PDF) of velocity circulation depends solely on the loop perimeter rather than its area. This perimeter rule holds strictly in the large-scale absolute equilibrium characterized by energy equipartition. At the intermediate states determined by both enstrophy and energy, these two regimes are separated by a characteristic equilibrium scale $l_\text{eq}$: the area rule governs loop statistics when $l\ll l_\text{eq}$, while the perimeter rule emerges for $l\gg l_\text{eq}$. These statistical laws remain robust even for loops with extreme aspect ratios as low as $0.03$, a value that inertial-range studies never achieved. Our findings provide a new steady-state solution to the 2D loop equation and suggest additional solutions, paving the way for exploring previously undiscovered geometric invariants of turbulence.

\end{abstract}

\maketitle

Velocity circulation, as an inviscid material invariant of the Navier-Stokes equation \cite{arnold1989,wu2015,arnold2021}, manifests itself in many perspectives of turbulence \cite{frisch1995,chen2006a}.
Inspired by the idea of Wilson loops \cite{wilson1974},  from the Navier-Stokes equation \citeauthor{migdal1994} \cite{migdal1994} derived a loop equation to describe the temporal evolution of the probability distribution function (PDF) of the velocity circulation over a fixed loop $C$.
Unlike the velocity structure function, which is commonly used to explore multiscale turbulence physics with scale defined as the two-point distance, the velocity circulation, $\varGamma_C= \oint_C\bm{u}\cdot\mathrm{d}\bm{x}$ with $\bm{u}$ the velocity field, depends on an infinite number of length scales determining the shape of loop $C$, such as perimeter, curvature, area, etc.
So far, the only steady solution to the loop equation is known as the area rule \cite{migdal2019a},
which states that the circulation PDF, $P(C,\varGamma,t)$,
depends solely on the minimal area of surfaces circumscribed by the loop $C$, and it is thought to be valid in the turbulent inertial range (IR) \cite{umeki1993,cao1996,iyer2019}.

According to the area rule, a universal scale for loops of different shapes can be defined as the square root of the minimal surface area circumscribed by the loops, enabling the study of the universal scaling of the velocity circulation over a range of scales.
Recent numerical and laboratory experiments reveal the universal bifractal behaviour of the velocity circulation moments over square loops in IR of both hydrodynamical \cite{iyer2019,zhu2023} and quantum \cite{muller2021,polanco2021,muller2024,massaro2025} turbulences: the scaling exponents of $p$-th order moment of velocity circulation, $\zeta_p\approx1.37p$ when $p<3$ and $\zeta_p\approx1.1p + 0.8$ when $p>3$.
Here, the low-order scaling $4p/3$ follows the \citeauthor{kolmogorov1941}'s IR theory \cite{kolmogorov1941},
and the high-order behaviour corresponds to a fractal set of dimension $2.2$ \cite{frisch1978,mandelbrot1974a}. 
Therefore, the velocity circulation is a potentially better quantity capturing the universality of turbulence compared with the velocity structure function, which is multifractal \cite{benzi1984,sreenivasan1997,sun2006,meneveau1987,she1994,yakhot2001,biferale2004} and non-universal \cite{chen1997,grossmann1997,boffetta2000}.

However, recent high-resolution direct numerical simulations (DNS) show that the area rule is not strictly valid in the IR of three-dimensional homogeneous isotropic turbulence (HIT) \cite{iyer2021} by demonstrating the aspect-ratio dependence of the velocity circulation moments over rectangular loops with the same minimal surface area.
Also, the area rule is incompatible with Kolmogorov's $ 5/3$ energy spectrum in the IR, considering the variance of velocity circulation \cite{xie2025}, which is linked to the energy spectrum through a Fourier transform.
Thus, two questions naturally arise: (i) Is there a realizable turbulent state in which the area rule holds?
When the area rule does not hold, (ii) are there any other turbulent statistically steady states capturing circulation's geometric invariants permitted by the loop equation beyond the circumscribed area?

In this letter, we show that the area rule is manifested in the large-scale dynamics of two-dimensional (2D) HIT, where the absolute equilibrium \cite{lee1951,kraichnan1967,kraichnan1975,gorce2022,dallas2015,kraichnan1973} with enstrophy equipartition is established.
Further, we derive from the 2D loop equation a \emph{perimeter rule}, which states that the circulation PDF depends solely on the loop perimeter rather than minimal surface area, and this perimeter rule holds strictly in the large-scale absolute equilibrium with energy equipartition of the  2D HIT

Based on a canonical ensemble with two quadratic conserved quantities, energy and enstrophy, of the truncated 2D Euler equation, Kraichnan \cite{kraichnan1967,kraichnan1975} derived the absolute-equilibrium energy and enstrophy spectra
\begin{equation}
    \label{eq:absolute equilibrium}
    E(k) = \frac{2\pi k}{\alpha + \beta k^2},\quad \Omega(k) = \frac{2\pi k^3}{\alpha+\beta k^2}.
\end{equation}
Here $k$ is the length of the 2D wavenumber $\bm{k}$, 
$\alpha$ and $\beta$ are Lagrangian multipliers corresponding to the conservations of energy and enstrophy, respectively.
In the analogy of Boson-Einstein condensation, $\alpha/\beta$ is the chemical potential \cite{kraichnan1975}, which corresponds to a characteristic wavenumber, $k_\text{eq}=\sqrt{\alpha/\beta}$, separating the energy and enstrophy equipartition states with spectra
\begin{equation}
\left\{
    \begin{aligned}
        E(k) \approx &2\pi k\alpha^{-1},& \Omega(k)\approx &2\pi k^{3}\alpha^{-1}, & k \ll k_\text{eq},\\
        E(k)\approx &2\pi k^{-1}\beta^{-1}, &\Omega(k)\approx &2\pi k\beta^{-1}, & k \gg k_\text{eq}.\\
    \end{aligned}
\right.
\end{equation}

For the truncated Euler system, a finite minimal scale of resolution, $\rho$, is fixed by the wavenumber of truncation, which imply us to introduced a spectral low-pass filter, $\exp(-\rho^2k^2/2)$. Therefore, for the energy-equipartition state, the velocity correlation reads,
\begin{equation}
\label{eq:correlation energy}
     \langle u_i(\bm{x})u_j(\bm{x}+\bm{r})\rangle = \frac{\delta_{ij} }{2\pi} \iint \frac{\mathrm{e}^{\mathrm{i}\bm{r}\cdot\bm{k}-\frac{\rho^2k^2}{2}}}{\alpha} \mathrm{d}k\mathrm{d}l = \frac{\delta_{ij}G_\rho(\bm{r})}{\alpha},
\end{equation}
and for the enstrophy-equipartition state, the vorticity correlation is,
\begin{equation}
\label{eq:correlation enstrophy}
\langle\omega(\bm{x})\omega(\bm{x}+\bm{r})\rangle = \frac{1}{2\pi}\iint \frac{\mathrm{e}^{\mathrm{i}\bm{r}\cdot\bm{k}-\frac{\rho^2k^2}{2}}}{\beta}\mathrm{d}k\mathrm{d}l = \frac{G_\rho(\bm{r})}{\beta}.
\end{equation}
Here, $G_\rho(\bm{r}) = \exp(-r^2\rho^{-2}/2)/(2\pi\rho^2)$ is the 2D Gaussian kernel with $\rho$ being a finite scale of resolution and $\delta_{ij}$ is the Kronecker's delta.

At the absolute equilibrium, the velocity and vorticity fields are both Gaussian random fields (GRF), thus, the velocity circulation obeys a Gaussian distribution determined by its variance $\langle\varGamma_C^2\rangle$, which is
\begin{equation}
\begin{aligned}
    \langle\varGamma_C^2\rangle =& \oint_C\oint_C \langle u_i(\bm{x})u_j(\bm{x}')\rangle\mathrm{d}x_i\mathrm{d}x_j'\\
    =&\iint_{S_C}\iint_{S_C} \langle\omega(\bm{x})\omega(\bm{x}')\rangle \mathrm{d}\sigma(\bm{x})\mathrm{d}\sigma(\bm{x}')
\end{aligned}
\end{equation}
Here and after, the summation over repeated indices is implied, $S_C$ represents the area enclosed by loop $C$ and $\mathrm{d}\sigma(\bm{x})$ represents the infinitesimal 2D area element at point $\bm{x}$.
Substitute the correlations (\ref{eq:correlation energy}) and (\ref{eq:correlation enstrophy}) into the expression of $\langle\varGamma_C^2\rangle$, we derive
\begin{equation}
     \begin{aligned}
    \langle\varGamma_C^2\rangle=\frac{L_{C,\rho}}{\sqrt{2\pi}\rho\alpha},\quad \langle\varGamma_C^2\rangle=\frac{A_{C,\rho}}{\beta}
    \end{aligned}
\end{equation}
for energy and enstrophy equipartition states, respectively.
Here, $L_{C,\rho}$ and $A_{C,\rho}$ are the coarse-grained perimeter and area of loop $C$, defined as,
\begin{equation}
\label{eq:reg peri}
    L_{C,\rho} = \sqrt{2\pi}\rho\oint_C\oint_C\delta_{ij}G_\rho(\bm{x}-\bm{x}')\mathrm{d}x_i\mathrm{d}x_j',
\end{equation}
and
\begin{equation}
    A_{C,\rho} = \iint_{S_C}\iint_{S_C}G_\rho(\bm{x}'-\bm{x}) \mathrm{d}\sigma(\bm{x})\mathrm{d}\sigma(\bm{x}'),
\end{equation}
which converge to the real area $A_C$ and perimeter $L_C$ for smooth loop $C$, respectively, when $\rho$ tends to zero.

\citeauthor{migdal2019d} showed that if $P(C,\varGamma,t)$ is scale-invariant and obeys the area rule, then
\begin{equation}
\label{eq:scale inv}
    P(C,\varGamma, t) = A_C^{-1/2} F(\varGamma A_C^{-1/2})
\end{equation}
with $F$ being an undetermined function\cite{migdal2019d}.
At the absolute equilibrium with enstrophy equipartition,
\begin{equation}
     P(C,\varGamma, t) = \sqrt{\frac{\beta}{2\pi A_{C,\rho}}}\exp\left(-\frac{\beta\varGamma^2}{2A_{C,\rho}}\right),
\end{equation}
which is consistent with the scale-invariant form (\ref{eq:scale inv}) with $F(X) = \sqrt{\beta/(2\pi)} \exp\left(-X^2/2\right)$ as the scale of resolution $\rho$ tends to zero. Thus, the absolute equilibrium and loop equation are consistent.

In the End Matter, we derive that the perimeter rule, 
\begin{equation}
\label{eq:perimeter rule}
    P(C,\varGamma,t) = F(L_{C,\rho},\varGamma),
\end{equation}
with $F$ an arbitrary smooth function, is a steady solution to the 2D inviscid loop equation,
\begin{equation}
    \label{eq:Two dimensional loop equation}
    \frac{\partial^2P(C,\varGamma,t)}{\partial \varGamma\partial t} = \oint_C\iint \frac{r_i}{2\pi r^2}\frac{\delta^2 P(C,\varGamma,t)}{\delta\sigma(\bm{x})\delta\sigma(\bm{x}')}\mathrm{d}\sigma(\bm{x}')\mathrm{d}x_i,
\end{equation}
where $r_i = x'_i-x_i$ and ${\delta}/{\delta\sigma(\bm{x})}$ is the 2D area derivative capturing the variation of the loop $C$. This loop equation (\ref{eq:Two dimensional loop equation}) is derived from the 2D Navier-Stokes equation,
\begin{equation}
\label{eq:2Dns}
    \frac{\partial\bm{u}}{\partial t} + \bm{u}\cdot\nabla\bm{u} = -\nabla p + \nu\nabla^2\bm{u},\quad \nabla\cdot\bm{u}=0.
\end{equation}

We can realize Kraichnan's absolute-equilibrium states in a forced-dissipative 2D HIT by carefully control the Reynolds number,
$Re := \varepsilon^{1/3}k_f^{-4/3}\nu^{-1}$, where $\varepsilon,k_f$ and $\nu$ are the energy-injection rate, forcing wavenumber and viscosity, respectively, and focus on the dynamics of scales larger than the forcing scale \cite{kan2025}. 
We employed the Julia package \texttt{GeophysicalFlows.jl}\cite{constantinou2021,constantinou2025}
to solve the 2D vorticity equation driven by a temporal white-noise force $f_\omega$ concentrated on wavenumber $k_f$ in a doubly periodic domain of size $2\pi L\times 2\pi L$
\begin{equation}
    \frac{\partial\omega}{\partial t} + \bm{u}\cdot\nabla\omega = f_\omega + \nu\nabla^2\omega.
\end{equation}
The $2/3$ rule is applied to eliminate aliasing errors, thus, the maximum resolved wavenumber is $k_\text{max} = N/3L$, where $N=1024$ is the number of grid points per domain side.
We simulated $7$ cases with $Re$ ranging from $3.5$ to $4.05$.
The forcing wavenumber of each run is designated as $k_f = k_\text{max}Re^{-1/2}$, such that the enstrophy dissipation scale, $(\varepsilon k_f^2)^{-1/6}\nu^{1/2} = k_\text{max}^{-1}$.
Even though the Reynolds numbers, which reflect the degrees of freedom between the scales of forcing and dissipation, are small in our simulations, the degrees of freedom in large scales, which are characterized by the ratio between the domain and forcing scales, are large.

\begin{figure}[htbp]
    \centering
    \includegraphics{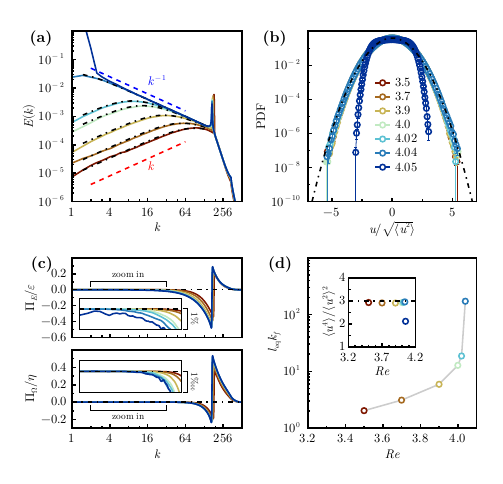}
    \caption{
        (a) Energy spectra for runs with different Reynolds numbers.
        The black dash-dotted lines illustrate the least square fitting results of the absolute equilibrium spectrum with positive $\alpha$ and $\beta$, and the red and blue dash-dotted lines denote the spectra corresponding to the energy and enstrophy equipartitions, $E(k)\propto k$ and $E(k)\propto k^{-1}$, respectively.
        (b) Normalized PDF of the large-scale velocity fluctuation for each run. The black dash-dotted line is the standard Gaussian distribution. 
        (c) Radial fluxes of energy (upper panel) and enstrophy (lower panel), normalized by the energy and enstrophy dissipation rates, respectively, and the insets show the zoom-in plots of the large-scale fluxes of energy and enstrophy.
        (d) $l_\text{eq}k_f$ and the kurtosis of the large-scale velocity fluctuation (inset) varying with $Re$.
        In all panels, $Re$ is encoded using the same colormap shown in the legend in (b).
    }
    \label{fig:diagnostic}
\end{figure}

Following \citeauthor{kan2025}, we verify the large-scale absolute equilibrium in our simulations with $Re$ ranging from $3.5$ to $4.04$.
FIG.\ \ref{fig:diagnostic} shows that, until a large-scale condensation emerges when $Re > 4.04$, (a) the large-scale energy spectra are well fitted by the absolute equilibrium spectrum (\ref{eq:absolute equilibrium}) with positive $\alpha$ and $\beta$, (b) the large-scale velocity fluctuations are approximately Gaussian and (c) no inverse cascade of energy is established.
In panel (d), we see the characteristic scale $l_\text{eq}$ diverges at $Re \approx 4.04$,
corresponding to the enstrophy equipartition at large scales.
The inset of panel (d) shows that the kurtosis of large-scale velocity fluctuations is close to $3$, reflecting the Gaussianity, until the large-scale condensation of energy emerges.

\begin{figure}[t]
    \centering
    \includegraphics{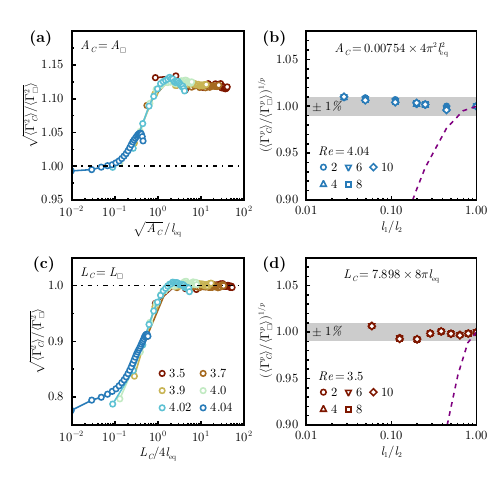}
    \caption{
        The ratios of variances of velocity circulations over rectangular loops $C$ with aspect ratio $4$ and square loops are plotted in (a) for identical areas $A_\square = A_C$, and plotted in (c) for identical perimeters $L_\square = L_C$.
        The loop sizes, $\sqrt{A_C}$ in (a) and $L_C/4$ in (c), are normalized by $ l_\text{eq}$ for runs with different $Re$.
        The ratios of moments of orders $2$ to $10$ for velocity circulation over rectangular and square loops are plotted with $Re = 4.04$ and fixed area $A_C$ in (b), plotted with $Re = 3.5$ and fixed perimeter $L_C$ in (d)
        The shadings show the region where the moment ratios deviate by less than $1\%$.
        The purple dash-dotted lines represent the corresponding variance ratios of velocity circulation for a synthetic 2D vorticity field with $\langle|\hat{\omega}|^2\rangle \propto k^{-2/3}$, corresponding to Kolmogorov's $-5/3$ law in IR.
        In all panels, the Reynolds numbers are encoded by the same colormap illustrated in the legend of panel (c).
    } 
    \label{fig:area perimeter}
\end{figure}

In FIG.\ \ref{fig:area perimeter}, we compare the velocity circulation over rectangular loops and square loops with identical areas (a,b) and perimeters (c,d). 
To focus on the large-scale dynamics, we filter out wavenumbers larger than $k_f$ when computing the velocity circulation.
By normalizing the loop sizes with $l_\text{eq}$, we see in panels (a) and (c) that the variance ratios of velocity circulation with different $l_\text{eq}$ collapse well,
and the area and perimeter rules hold for loop sizes much smaller and larger than $l_\text{eq}$, respectively.
In panels (b) and (d), we see that the higher-order moments of velocity circulation still obey the area or perimeter rule even when the aspect ratio reaches $0.03$, an extreme value that previous studies \cite[cf.][]{iyer2021} did not reach,
and the deviations are less than $1\%$, which again justifies the statistically equilibrium states.
Further, the velocity circulation in IR (at least its variance) obeys neither the area nor the perimeter rule in the range of aspect ratios considered in this letter, as illustrated by the purple dashed lines in panels (b) and (d).

\begin{figure}[htbp]
    \centering
    \includegraphics{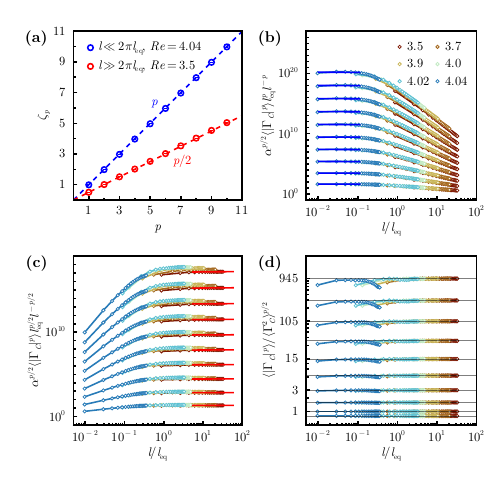}
    \caption{
        (a) Scaling exponents $\zeta_p$ of the $p$-th order moments of velocity circulation for loop sizes $l\ll l_\text{eq}$ (blue markers, fitted for $Re = 4.04$) and $l\gg l_\text{eq}$ (red markers, fitted for $Re = 3.5$)
        The dashed lines refer to $\zeta_p = p$ (blue) and $\zeta_p =  p/2$ (red line), respectively. 
        The $p$-th order moments of the velocity circulation $\langle|\varGamma_C|^p\rangle$ compensated by $l^{-p}$ (b) and $l^{-p/2}$ (c),
        where the veolcity circulations are normalized by $\alpha^{1/2}$ and the loop sizes $l$ are normalized by $l_\text{eq}$.
        The blue and red lines in (b) and (c) are the least-square fitting results corresponding to each scaling exponent in panel (a).
        (d) Normalized $p$-th order moments of velocity circulation. The horizontal lines illustrate the $p$-th order moments for the standard Gaussian distribution.
    }
    \label{fig:scaling exponents}
\end{figure}

In FIG.\ \ref{fig:scaling exponents}, we see that the scaling exponents $\zeta_p$,
where $\langle|\varGamma_C|^p\rangle \propto l^{\zeta_p}$ with $l$ being the loop size,
show no intermittency and are fitted well by $\zeta_p =p$ for $l\ll l_\text{eq}$ and $\zeta_p = p/2$ for $l\gg l_\text{eq}$, corresponding to $\langle \varGamma_C^2\rangle$ being proportional to the loop area $A_C$ and perimeter $L_C$, respectively. In panel (d), we compare the normalized $p$-th order moments of velocity circulation for $Re$ ranging from $3.5$ to $4.04$ with that of the standard Gaussian distribution,
which shows that the velocity circulation PDFs are very close to the Gaussian distribution, providing evidence that the large-scale velocity and vorticity fields are GRFs. 

To demonstrate the convergence of the statistical moments, we plot the integrands of the $10$th-order moments of velocity circulations, $\varGamma_C^{10}P(\varGamma_C)$, in section S2 of the supplementary material.

In summary, we have shown that velocity circulation in the large scales of 2D turbulence obeys the area and perimeter rules when the absolute equilibrium is established and the loop size $l\ll l_\text{eq}$ and $l\gg l_\text{eq}$, respectively, even for rectangles with very extreme aspect ratios.
And theoretically, we demonstrate that the perimeter rule is a new steady solution to the 2D inviscid loop equation (\ref{eq:Two dimensional loop equation}), implying that there exist other quantities dominating the statistics of the velocity circulation beyond the loop area.
In our previous discussion, the area rule of the circulation variance could be derived from the enstrophy spectrum, $\langle|\hat{\omega}|^2\rangle(\bm{k}) = \beta^{-1}$.
Analogically, the enstrophy spectrum in the IR, $\langle|\hat{\omega}|^2\rangle(\bm{k})\propto k^{-2/3}$, should correspond to a new geometrical quantity about loop $C$,
\begin{equation}
    X_C \propto \iint_{S_C}\iint_{S_C} (-\nabla^{2})^{-1/3}\delta(\bm{x}-\bm{x}')\mathrm{d}\sigma(\bm{x})\mathrm{d}\sigma(\bm{x}'),
\end{equation}
which may capture certain fractal structures within the turbulent flow in the IR, and its characteristic geometry is neither surfaces nor lines.
Further, we noticed that the dependency of circulation statistics on loop perimeters was observed in quantum turbulence \cite{massaro2025}, where the vortex-antivortex pairs dominate the dynamics.
We expect that the perimeter rule could be manifested in point vortex systems\cite{lundgren1977,gauthier2019} as well, where the vortices with opposite signs tend to pair up when the Hamiltonian of the point vortex system tends to $-\infty$.
For the absolute equilibrium state with large-scale condensation, \citeauthor{agoua2025}\cite{agoua2025} shows that the flow field can be divided into a condensate part and an incoherent part with enstrophy equipartitioned, which corresponds to the area rule of velocity circulation.
The perimeter rule is also expected to hold in the large-scale 3D HIT, where the absolute equilibrium is established and exhibits a Rayleigh-Jeans spectrum\cite{dallas2015,gorce2022}, $E(k)\propto k^2$, characterizing an equipartition of energy in 3D HIT.

\begin{acknowledgments}
We acknowledge financial support from the National Natural Science Foundation of China, grant numbers 12272006, 12472219, 42361144844 and 12588201, and from the Laoshan Laboratory under grant numbers LSKJ202202000, LSKJ202300100, LSJKJ202400203.
\end{acknowledgments}

\bibliographystyle{apsrev4-2}
\bibliography{ref}

@article{agoua2025,
  title = {Coexistence of Two Equilibrium Configurations in Two-Dimensional Turbulence},
  author = {Agoua, Wesley and Yin, Xi-Yuan and Wu, Tong and Bos, Wouter J. T.},
  year = 2025,
  month = mar,
  journal = {Physical Review Fluids},
  volume = {10},
  number = {3},
  pages = {034604},
  publisher = {American Physical Society},
  doi = {10.1103/PhysRevFluids.10.034604},
  urldate = {2025-09-06}
}

@book{arnold1989,
  title = {Mathematical {{Methods Of Classical Mechanics}}},
  author = {Arnold, V. I. and Weinstein, A. and Vogtmann, K.},
  year = 1989,
  series = {Graduate {{Texts}} in {{Mathematics}}},
  edition = {2nd},
  publisher = {Springer},
  urldate = {2022-05-27},
  isbn = {978-0-387-96890-2}
}

@book{arnold2021,
  title = {Topological {{Methods}} in {{Hydrodynamics}}},
  author = {Arnold, Vladimir I. and Khesin, Boris A.},
  year = 2021,
  series = {Applied {{Mathematical Sciences}}},
  volume = {125},
  publisher = {Springer International Publishing},
  address = {Cham},
  doi = {10.1007/978-3-030-74278-2},
  urldate = {2025-12-21},
  copyright = {https://www.springer.com/tdm},
  isbn = {978-3-030-74277-5 978-3-030-74278-2},
  langid = {english}
}

@article{benzi1984,
  title = {On the Multifractal Nature of Fully Developed Turbulence and Chaotic Systems},
  author = {Benzi, R. and Paladin, G. and Parisi, G. and Vulpiani, A.},
  year = 1984,
  month = dec,
  journal = {Journal of Physics A: Mathematical and General},
  volume = {17},
  number = {18},
  pages = {3521},
  issn = {0305-4470},
  doi = {10.1088/0305-4470/17/18/021},
  urldate = {2025-06-22},
  langid = {english}
}

@article{biferale2004,
  title = {Multifractal {{Statistics}} of {{Lagrangian Velocity}} and {{Acceleration}} in {{Turbulence}}},
  author = {Biferale, L. and Boffetta, G. and Celani, A. and Devenish, B. J. and Lanotte, A. and Toschi, F.},
  year = 2004,
  month = aug,
  journal = {Physical Review Letters},
  volume = {93},
  number = {6},
  pages = {064502},
  issn = {0031-9007, 1079-7114},
  doi = {10.1103/PhysRevLett.93.064502},
  urldate = {2025-11-17},
  copyright = {http://link.aps.org/licenses/aps-default-license},
  langid = {english}
}

@article{boffetta2000,
  title = {Inverse Energy Cascade in Two-Dimensional Turbulence: {{Deviations}} from {{Gaussian}} Behavior},
  shorttitle = {Inverse Energy Cascade in Two-Dimensional Turbulence},
  author = {Boffetta, G. and Celani, A. and Vergassola, M.},
  year = 2000,
  month = jan,
  journal = {Physical Review E},
  volume = {61},
  number = {1},
  pages = {R29-R32},
  publisher = {American Physical Society},
  doi = {10.1103/PhysRevE.61.R29},
  urldate = {2025-09-02},
  langid = {american}
}

@article{cao1996,
  title = {Properties of {{Velocity Circulation}} in {{Three-Dimensional Turbulence}}},
  author = {Cao, Nianzheng and Chen, Shiyi and Sreenivasan, Katepalli R.},
  year = 1996,
  month = jan,
  journal = {Physical Review Letters},
  volume = {76},
  number = {4},
  pages = {616--619},
  publisher = {American Physical Society},
  doi = {10.1103/PhysRevLett.76.616},
  urldate = {2025-09-06}
}

@article{chen1997,
  title = {Refined {{Similarity Hypothesis}} for {{Transverse Structure Functions}} in {{Fluid Turbulence}}},
  author = {Chen, Shiyi and Sreenivasan, Katepalli R. and Nelkin, Mark and Cao, Nianzheng},
  year = 1997,
  month = sep,
  journal = {Physical Review Letters},
  volume = {79},
  number = {12},
  pages = {2253--2256},
  publisher = {American Physical Society},
  doi = {10.1103/PhysRevLett.79.2253},
  urldate = {2025-10-26}
}

@article{chen2006a,
  title = {Is the {{Kelvin Theorem Valid}} for {{High Reynolds Number Turbulence}}?},
  author = {Chen, Shiyi and Eyink, Gregory L. and Wan, Minping and Xiao, Zuoli},
  year = 2006,
  month = oct,
  journal = {Physical Review Letters},
  volume = {97},
  number = {14},
  pages = {144505},
  publisher = {American Physical Society},
  doi = {10.1103/PhysRevLett.97.144505},
  urldate = {2025-11-27}
}

@article{constantinou2021,
  title = {{{GeophysicalFlows}}.Jl: {{Solvers}} for Geophysical Fluid Dynamics Problems in Periodic Domains on {{CPUs}} \& {{GPUs}}},
  shorttitle = {{{GeophysicalFlows}}.Jl},
  author = {Constantinou, Navid C. and Wagner, Gregory LeClaire and Siegelman, Lia and Pearson, Brodie C. and Pal{\'o}czy, Andr{\'e}},
  year = 2021,
  month = apr,
  journal = {Journal of Open Source Software},
  volume = {6},
  number = {60},
  pages = {3053},
  issn = {2475-9066},
  doi = {10.21105/joss.03053},
  urldate = {2025-09-17},
  langid = {english}
}

@misc{constantinou2025,
  title = {{{FourierFlows}}/{{FourierFlows}}.Jl: V0.10.7},
  shorttitle = {{{FourierFlows}}/{{FourierFlows}}.Jl},
  author = {Constantinou, Navid C. and Wagner, Gregory L. and Pal{\'o}czy, Andr{\'e} and HO, Ka Wai and Bisits, Josef and TagBot, Julia and Piibeleht, Morten and Besard, Tim and Robertson, Connor and Parfenyev, Vladimir},
  year = 2025,
  month = oct,
  doi = {10.5281/zenodo.17281674},
  urldate = {2025-12-21},
  howpublished = {Zenodo}
}

@article{dallas2015,
  title = {Statistical {{Equilibria}} of {{Large Scales}} in {{Dissipative Hydrodynamic Turbulence}}},
  author = {Dallas, V. and Fauve, S. and Alexakis, A.},
  year = 2015,
  month = nov,
  journal = {Physical Review Letters},
  volume = {115},
  number = {20},
  pages = {204501},
  issn = {0031-9007, 1079-7114},
  doi = {10.1103/PhysRevLett.115.204501},
  urldate = {2025-05-15},
  copyright = {http://link.aps.org/licenses/aps-default-license},
  langid = {english}
}

@book{frisch1995,
  title = {Turbulence},
  author = {Frisch, Uriel},
  year = 1995,
  edition = {Paperback},
  publisher = {Cambridge University Press},
  urldate = {2023-02-10},
  isbn = {978-0-521-45713-2}
}

@article{gauthier2019,
  title = {Giant Vortex Clusters in a Two-Dimensional Quantum Fluid},
  author = {Gauthier, Guillaume and Reeves, Matthew T. and Yu, Xiaoquan and Bradley, Ashton S. and Baker, Mark A. and Bell, Thomas A. and {Rubinsztein-Dunlop}, Halina and Davis, Matthew J. and Neely, Tyler W.},
  year = 2019,
  month = jun,
  journal = {Science},
  volume = {364},
  number = {6447},
  pages = {1264--1267},
  publisher = {American Association for the Advancement of Science},
  doi = {10.1126/science.aat5718},
  urldate = {2025-03-31},
  langid = {american}
}

@article{gorce2022,
  title = {Statistical {{Equilibrium}} of {{Large Scales}} in {{Three-Dimensional Hydrodynamic Turbulence}}},
  author = {Gorce, Jean-Baptiste and Falcon, Eric},
  year = 2022,
  month = jul,
  journal = {Physical Review Letters},
  volume = {129},
  number = {5},
  pages = {054501},
  issn = {0031-9007, 1079-7114},
  doi = {10.1103/PhysRevLett.129.054501},
  urldate = {2025-05-15},
  langid = {english}
}

@article{grossmann1997,
  title = {Different Intermittency for Longitudinal and Transversal Turbulent Fluctuations},
  author = {Grossmann, Siegfried and Lohse, Detlef and Reeh, Achim},
  year = 1997,
  month = dec,
  journal = {Physics of Fluids},
  volume = {9},
  number = {12},
  pages = {3817--3825},
  issn = {1070-6631},
  doi = {10.1063/1.869516},
  urldate = {2025-10-26}
}

@article{iyer2019,
  title = {Circulation in {{High Reynolds Number Isotropic Turbulence}} Is a {{Bifractal}}},
  author = {Iyer, Kartik P. and Sreenivasan, Katepalli R. and Yeung, P. K.},
  year = 2019,
  month = oct,
  journal = {Physical Review X},
  volume = {9},
  number = {4},
  pages = {041006},
  issn = {2160-3308},
  doi = {10.1103/PhysRevX.9.041006},
  urldate = {2025-03-11},
  langid = {english}
}

@article{iyer2021,
  title = {The Area Rule for Circulation in Three-Dimensional Turbulence},
  author = {Iyer, Kartik P. and Bharadwaj, Sachin S. and Sreenivasan, Katepalli R.},
  year = 2021,
  month = oct,
  journal = {Proceedings of the National Academy of Sciences},
  volume = {118},
  number = {43},
  pages = {e2114679118},
  issn = {0027-8424, 1091-6490},
  doi = {10.1073/pnas.2114679118},
  urldate = {2025-03-11},
  langid = {english}
}

@misc{kan2025,
  title = {Two-Dimensional Turbulent Condensates without Bottom Drag},
  author = {van Kan, Adrian and Alexakis, Alexandros and Knobloch, Edgar},
  year = 2025,
  month = apr,
  number = {arXiv:2504.02978},
  eprint = {2504.02978},
  primaryclass = {physics},
  publisher = {arXiv},
  doi = {10.48550/arXiv.2504.02978},
  urldate = {2025-06-11},
  archiveprefix = {arXiv},
  langid = {american}
}

@article{kolmogorov1941,
  title = {The {{Local Structure}} of {{Turbulence}} in {{Incompressible Viscous Fluid}} for {{Very Large Reynolds}}' {{Numbers}}},
  author = {Kolmogorov, A.},
  year = 1941,
  month = jan,
  journal = {Akademiia Nauk SSSR Doklady},
  volume = {30},
  pages = {301--305},
  issn = {0002-3264},
  urldate = {2025-11-17}
}

@article{kraichnan1967,
  title = {Inertial {{Ranges}} in {{Two}}-{{Dimensional Turbulence}}},
  author = {Kraichnan, Robert H.},
  year = 1967,
  month = jul,
  journal = {The Physics of Fluids},
  volume = {10},
  number = {7},
  pages = {1417--1423},
  issn = {0031-9171},
  doi = {10.1063/1.1762301},
  urldate = {2025-11-17}
}

@article{kraichnan1973,
  title = {Helical Turbulence and Absolute Equilibrium},
  author = {Kraichnan, Robert H.},
  year = 1973,
  month = aug,
  journal = {Journal of Fluid Mechanics},
  volume = {59},
  number = {4},
  pages = {745--752},
  issn = {1469-7645, 0022-1120},
  doi = {10.1017/S0022112073001837},
  urldate = {2025-06-25},
  langid = {english}
}

@article{kraichnan1975,
  title = {Statistical Dynamics of Two-Dimensional Flow},
  author = {Kraichnan, Robert H.},
  year = 1975,
  month = jan,
  journal = {Journal of Fluid Mechanics},
  volume = {67},
  number = {1},
  pages = {155--175},
  issn = {1469-7645, 0022-1120},
  doi = {10.1017/S0022112075000225},
  urldate = {2025-09-06},
  langid = {english}
}

@article{lee1951,
  title = {Difference between {{Turbulence}} in a {{Two}}-{{Dimensional Fluid}} and in a {{Three}}-{{Dimensional Fluid}}},
  author = {Lee, T. D.},
  year = 1951,
  month = apr,
  journal = {Journal of Applied Physics},
  volume = {22},
  number = {4},
  pages = {524},
  issn = {0021-8979},
  doi = {10.1063/1.1699997},
  urldate = {2025-11-17}
}

@article{lundgren1977,
  title = {Statistical Mechanics of Two-Dimensional Vortices},
  author = {Lundgren, T. S. and Pointin, Y. B.},
  year = 1977,
  month = nov,
  journal = {Journal of Statistical Physics},
  volume = {17},
  number = {5},
  pages = {323--355},
  issn = {1572-9613},
  doi = {10.1007/BF01014402},
  urldate = {2025-02-16},
  langid = {english}
}

@misc{massaro2025,
  title = {Circulation {{Statistics}} and {{Migdal Area Rule Beyond}} the {{Kibble-Zurek Mechanism}} in a {{Newborn Bose-Einstein Condensate}}},
  author = {Massaro, Matteo and Shinn, Seong-Ho and Thudiyangal, Mithun and del Campo, Adolfo},
  year = 2025,
  month = aug,
  number = {arXiv:2508.11047},
  eprint = {2508.11047},
  primaryclass = {cond-mat},
  publisher = {arXiv},
  doi = {10.48550/arXiv.2508.11047},
  urldate = {2025-08-19},
  archiveprefix = {arXiv},
  langid = {english}
}

@article{meneveau1987,
  title = {Simple Multifractal Cascade Model for Fully Developed Turbulence},
  author = {Meneveau, C. and Sreenivasan, K. R.},
  year = 1987,
  month = sep,
  journal = {Physical Review Letters},
  volume = {59},
  number = {13},
  pages = {1424--1427},
  issn = {0031-9007},
  doi = {10.1103/PhysRevLett.59.1424},
  urldate = {2025-11-17},
  copyright = {http://link.aps.org/licenses/aps-default-license},
  langid = {english}
}

@article{migdal1994,
  title = {Loop {{Equation}} and {{Area Law}} in {{Turbulence}}},
  author = {Migdal, Alexander A.},
  year = 1994,
  month = mar,
  journal = {International Journal of Modern Physics A},
  volume = {09},
  number = {08},
  eprint = {hep-th/9310088},
  pages = {1197--1238},
  issn = {0217-751X, 1793-656X},
  doi = {10.1142/S0217751X94000558},
  urldate = {2025-03-19},
  archiveprefix = {arXiv}
}

@misc{migdal2019a,
  title = {Universal {{Area Law}} in {{Turbulence}}},
  author = {Migdal, Alexander},
  year = 2019,
  month = apr,
  number = {arXiv:1903.08613},
  eprint = {1903.08613},
  primaryclass = {hep-th},
  publisher = {arXiv},
  doi = {10.48550/arXiv.1903.08613},
  urldate = {2025-03-11},
  archiveprefix = {arXiv},
  langid = {american}
}

@misc{migdal2019d,
  title = {Scaling {{Index}} {$\alpha={1}/{2}$} {{In Turbulent Area Law}}},
  author = {Migdal, Alexander},
  year = 2019,
  month = apr,
  number = {arXiv:1904.00900},
  eprint = {1904.00900},
  primaryclass = {hep-th},
  publisher = {arXiv},
  doi = {10.48550/arXiv.1904.00900},
  urldate = {2025-12-23},
  archiveprefix = {arXiv}
}

@article{muller2021,
  title = {Intermittency of {{Velocity Circulation}} in {{Quantum Turbulence}}},
  author = {M{\"u}ller, Nicol{\'a}s P. and Polanco, Juan Ignacio and Krstulovic, Giorgio},
  year = 2021,
  month = mar,
  journal = {Physical Review X},
  volume = {11},
  number = {1},
  pages = {011053},
  publisher = {American Physical Society},
  doi = {10.1103/PhysRevX.11.011053},
  urldate = {2025-03-19}
}

@article{muller2024,
  title = {Exploring the {{Equivalence}} between {{Two-Dimensional Classical}} and {{Quantum Turbulence}} through {{Velocity Circulation Statistics}}},
  author = {M{\"u}ller, Nicol{\'a}s P. and Krstulovic, Giorgio},
  year = 2024,
  month = feb,
  journal = {Physical Review Letters},
  volume = {132},
  number = {9},
  pages = {094002},
  publisher = {American Physical Society},
  doi = {10.1103/PhysRevLett.132.094002},
  urldate = {2025-09-06}
}

@article{polanco2021,
  title = {Vortex Clustering, Polarisation and Circulation Intermittency in Classical and Quantum Turbulence},
  author = {Polanco, Juan Ignacio and M{\"u}ller, Nicol{\'a}s P. and Krstulovic, Giorgio},
  year = 2021,
  month = dec,
  journal = {Nature Communications},
  volume = {12},
  number = {1},
  pages = {7090},
  issn = {2041-1723},
  doi = {10.1038/s41467-021-27382-6},
  urldate = {2025-05-20},
  langid = {english}
}

@article{she1994,
  title = {Universal Scaling Laws in Fully Developed Turbulence},
  author = {She, Zhen-Su and Leveque, Emmanuel},
  year = 1994,
  month = jan,
  journal = {Physical Review Letters},
  volume = {72},
  number = {3},
  pages = {336--339},
  publisher = {American Physical Society},
  doi = {10.1103/PhysRevLett.72.336},
  urldate = {2025-07-01},
  langid = {american}
}

@article{sreenivasan1997,
  title = {{{THE PHENOMENOLOGY OF SMALL-SCALE TURBULENCE}}},
  author = {Sreenivasan, K. R. and Antonia, R. A.},
  year = 1997,
  month = jan,
  journal = {Annual Review of Fluid Mechanics},
  volume = {29},
  number = {Volume 29, 1997},
  pages = {435--472},
  publisher = {Annual Reviews},
  issn = {0066-4189, 1545-4479},
  doi = {10.1146/annurev.fluid.29.1.435},
  urldate = {2025-07-01},
  langid = {english}
}

@article{sun2006,
  title = {Cascades of {{Velocity}} and {{Temperature Fluctuations}} in {{Buoyancy-Driven Thermal Turbulence}}},
  author = {Sun, Chao and Quan, Zhou and {Ke-Qing}, Xia},
  year = 2006,
  journal = {Physical Review Letters},
  volume = {97},
  number = {14},
  doi = {10.1103/PhysRevLett.97.144504}
}

@article{umeki1993,
  title = {Probability {{Distribution}} of {{Velocity Circulation}}   in {{Three-Dimensional Turbulence}}},
  author = {Umeki, Makoto},
  year = 1993,
  month = nov,
  journal = {Journal of the Physical Society of Japan},
  volume = {62},
  number = {11},
  pages = {3788--3791},
  publisher = {The Physical Society of Japan},
  issn = {0031-9015},
  doi = {10.1143/JPSJ.62.3788},
  urldate = {2025-04-07}
}

@book{wu2015,
  title = {Vortical {{Flows}}},
  author = {Wu, Jie-Zhi and Ma, Hui-Yang and Zhou, Ming-De},
  year = 2015,
  publisher = {Springer Berlin Heidelberg},
  address = {Berlin, Heidelberg},
  doi = {10.1007/978-3-662-47061-9},
  urldate = {2024-12-05},
  copyright = {https://www.springernature.com/gp/researchers/text-and-data-mining},
  isbn = {978-3-662-47060-2 978-3-662-47061-9},
  langid = {english}
}

@misc{xie2025,
  title = {Area Rule of Velocity Circulation in Two-Dimensional Instability-Driven Turbulence beyond the Inertial Range},
  author = {Xie, Bo-Jie and Xie, Jin-Han},
  year = 2025,
  month = apr,
  number = {arXiv:2504.21512},
  eprint = {2504.21512},
  primaryclass = {physics},
  publisher = {arXiv},
  doi = {10.48550/arXiv.2504.21512},
  urldate = {2025-09-09},
  archiveprefix = {arXiv}
}

@article{yakhot2001,
  title = {Mean-Field Approximation and a Small Parameter in Turbulence Theory},
  author = {Yakhot, Victor},
  year = 2001,
  month = jan,
  journal = {Physical Review E},
  volume = {63},
  number = {2},
  pages = {026307},
  publisher = {American Physical Society},
  doi = {10.1103/PhysRevE.63.026307},
  urldate = {2025-11-17}
}

@article{zhu2023,
  title = {Circulation in {{Quasi-2D Turbulence}}: {{Experimental Observation}} of the {{Area Rule}} and {{Bifractality}}},
  shorttitle = {Circulation in {{Quasi-2D Turbulence}}},
  author = {Zhu, Hang-Yu and Xie, Jin-Han and Xia, Ke-Qing},
  year = 2023,
  month = may,
  journal = {Physical Review Letters},
  volume = {130},
  number = {21},
  pages = {214001},
  issn = {0031-9007, 1079-7114},
  doi = {10.1103/PhysRevLett.130.214001},
  urldate = {2025-03-19},
  langid = {english}
}

@article{wilson1974,
  title = {Confinement of Quarks},
  author = {Wilson, Kenneth G.},
  year = 1974,
  journal = {Physical Review D},
  volume = {10},
  number = {8},
  pages = {2445--2459},
  doi = {10.1103/PhysRevD.10.2445}
}

@article{frisch1978,
  title = {A Simple Dynamical Model of Intermittent Fully Developed Turbulence},
  author = {Frisch, Uriel and Sulem, Pierre-Louis and Nelkin, Mark},
  year = 1978,
  month = aug,
  journal = {Journal of Fluid Mechanics},
  volume = {87},
  number = {4},
  pages = {719--736},
  issn = {1469-7645, 0022-1120},
  doi = {10.1017/S0022112078001846},
  urldate = {2026-01-05},
  langid = {english}
}

@article{mandelbrot1974a,
  title = {Intermittent Turbulence in Self-Similar Cascades: Divergence of High Moments and Dimension of the Carrier},
  shorttitle = {Intermittent Turbulence in Self-Similar Cascades},
  author = {Mandelbrot, Benoit B.},
  year = 1974,
  month = jan,
  journal = {Journal of Fluid Mechanics},
  volume = {62},
  number = {2},
  pages = {331--358},
  issn = {1469-7645, 0022-1120},
  doi = {10.1017/S0022112074000711},
  urldate = {2026-01-21},
  langid = {english}
}

\clearpage

\begin{center}
    \textbf{END MATTER}
\end{center}

\section{2D Loop Equation}
For a fixed Eulerian loop $C$ in 2D space, the PDF of velocity circulation can be expressed as
\begin{equation}
    P(C,\varGamma,t) = \left\langle \delta\left(\varGamma - \oint_C \bm{u}\cdot\mathrm{d}\bm{x}\right)\right\rangle,
\end{equation}
whose partial derivatives give
\begin{equation}
\label{eq:partial derivative}
    \frac{\partial^2P}{\partial\varGamma\partial t} = -\left\langle \delta''\left( \varGamma -\oint_C\bm{u}\cdot\mathrm{d}\bm{x}\right)\oint_C\frac{\partial\bm{u}}{\partial t}\cdot\mathrm{d}\bm{x}\right\rangle.
\end{equation}
Substitute $\partial_t\bm{u}$ with the Lamb form of 2D Navier-Stokes equation(\ref{eq:2Dns})
and express the velocity field with the Biot-Savart theorem, we have
\begin{equation}
\label{eq:circulation evolv}
    \begin{aligned}
        \oint_C\frac{\partial \bm{u}}{\partial t}\cdot\mathrm{d}\bm{x} =&\oint_C\left(\iint\frac{r_i\omega\omega'}{2\pi r^2}\mathrm{d}\sigma' - \epsilon_{ij}\nu\partial_j\omega \right)\mathrm{d}x_i \\
    \end{aligned}
\end{equation}
where $\epsilon_{ij}$ is the 2D Levi-Civita tensor, $r_i = x'_i-x_i$, $\omega'$ and $\mathrm{d}\sigma'$ are the vorticity and infinitesimal area element at point $\bm{x}'$.

In 2D, we define the area derivative for an arbitrary functional $F(C)$ about loop $C$ at point $\bm{x}\in\mathbb{R}^2$ as
\begin{equation}
\label{eq:area derivative}
    \frac{\delta F(C)}{\delta\sigma(\bm{x})} := \lim_{\delta C\to C_{\bm{x}}} \frac{F(C+\delta C) - F(C)}{a_{\delta C}},
\end{equation}
where the limit $\delta C \to C_{\bm{x}}$ means that the loop $\delta C$ shrinks to circumscribe a single point $\bm{x}\in\mathbb{R}^2$, $a_C$ is the linear oriented area for a loop $C$,
\begin{equation}
    a_C = \frac{1}{2}\epsilon_{ij}\oint_C x_i\mathrm{d}x_j,
\end{equation}
and the addition of loops, $C + \delta C$, represents a new loop constructed by connecting $C$ and $\delta C$ through a pair of lines with opposite direction.

By definition, the area derivative of the velocity circulation gives the vorticity at $\bm{x}$,
\begin{equation}
    \frac{\delta \left(\oint_Cu_i\mathrm{d}x_i\right)}{\delta\sigma(\bm{x})} = \lim_{\delta C\to C_{\bm{x}}}a_{\delta C}^{-1} \oint_{\delta C}u_i\mathrm{d}x_i= \omega(\bm{x}),
\end{equation}
thus, we could obtain the vorticity at multiple points by applying area derivatives at corresponding locations to $P(C,\varGamma,t)$, e.g.
\begin{equation}
\label{eq: second derivative}
    \frac{\delta^2 P(C,\varGamma,t)}{\delta\sigma(\bm{x})\delta\sigma(\bm{x}')} = \left\langle \omega(\bm{x})\omega(\bm{x}')\delta''\left( \varGamma -\oint_C \bm{u}\cdot\mathrm{d}\bm{x}\right) \right\rangle.
\end{equation}
Comparing (\ref{eq:partial derivative}), (\ref{eq:circulation evolv}) and (\ref{eq: second derivative}), we could derive a closed equation about $P(C,\varGamma,t)$,
\begin{widetext}
\begin{equation}
    \frac{\partial^2P(C,\varGamma,t)}{\partial \varGamma\partial t} =\oint_C\left(\iint \frac{r_i}{2\pi r^2}\frac{\delta^2 P(C,\varGamma,t)}{\delta\sigma(\bm{x})\delta\sigma(\bm{x}')}\mathrm{d}\sigma' + \nu\epsilon_{ij}\frac{\partial^2}{\partial\varGamma\partial x_j}\frac{\delta P(C,\varGamma,t)}{\delta\sigma(\bm{x})}\right)\mathrm{d}x_i,
\end{equation}
\end{widetext}
and at the inviscid limit, $\nu\to 0$, we have the inviscid 2D loop equation,
\begin{equation}
    \frac{\partial^2P(C,\varGamma,t)}{\partial \varGamma\partial t}=\oint_C\iint \frac{r_i}{2\pi r^2}\frac{\delta^2 P(C,\varGamma,t)}{\delta\sigma(\bm{x})\delta\sigma(\bm{x}')}\mathrm{d}\sigma'\mathrm{d}x_i.
\end{equation}

\section{Regularized Loop Perimeter}

We show that, the coarse-grained perimeter (\ref{eq:reg peri}), $L_{C,\rho}$, converges to the real perimeter $L_C$ when $\rho\to 0$ for a smooth non-self-intersecting loop $C$.
A smooth loop $C$ can be locally expanded at point $\bm{x}$ as
\begin{equation}
\label{eq:curve local expansion}
    \bm{x}'-\bm{x} = \bm{t}\tilde{s} + \frac{1}{2}\kappa\bm{n}\tilde{s}^2 + \frac{1}{6}(\kappa_s\bm{n}-\kappa^2\bm{t})\tilde{s}^3 + \text{h.o.t.},
\end{equation}
where $\text{h.o.t.}$ means higher order terms, $\tilde{s}=s'-s$ with $s$ and $s'$ being the arc-length coordinates at points $\bm{x}$ and $\bm{x}'$ on $C$, respectively,
$\bm{t}$, $\bm{n}$, $\kappa$ and $\kappa_s$ are the tangent vector, normal vector, curvature and derivative of curvature at $\bm{x}$.
Substitute the local expansion (\ref{eq:curve local expansion}) into $L_{C,\rho}$ (\ref{eq:reg peri}), we could derive an asymptotic series,
\begin{equation}
    \begin{aligned}
    L_{C,\rho} =&\oint_C\oint_C\left(1-\frac{\kappa^2\tilde{s}^2}{2} +\text{h.o.t.}\right)\frac{\mathrm{e}^{-\frac{\tilde{s}^2 + O(\tilde{s}^4)}{2\rho^2}}}{\sqrt{2\pi}\rho}\mathrm{d}s'\mathrm{d}s\\
   =& \oint_C1\mathrm{d}s + \frac{\rho^2}{2}\oint_C k^2 \mathrm{d}s + \text{h.o.t.},
    \end{aligned}
\end{equation}
whose leading term is exactly the loop perimeter $L_C$.

\section{Perimeter Rule}

To show that the perimeter rule is a steady solution of the 2D inviscid loop equation, we substitute the coarse-grained loop perimeter (\ref{eq:perimeter rule}), $L_{C,\rho}$, into the 2D inviscid loop equation (\ref{eq:Two dimensional loop equation}) and show that the right-hand-side vanishes for arbitrary smooth loop $C$. The substitution gives
\begin{equation}
    \frac{\partial^2 P(C,\varGamma,t)}{\partial \varGamma\partial t} = \frac{\mathcal{I}_{C,\rho}}{2\pi}\frac{\partial F}{\partial L_{C,\rho}} + \frac{\mathcal{J}_{C,\rho}}{2\pi}\frac{\partial^2 F}{\partial L_{C,\rho}^2},
\end{equation}
where
\begin{align}
    \mathcal{I}_{C,\rho} = & \oint_C\iint\frac{r_i}{r^2}\frac{\delta^2 L_{C,\rho}}{\delta\sigma(\bm{x})\delta\sigma(\bm{x}')}\mathrm{d}\sigma'\mathrm{d}x_i,\\
    \mathcal{J}_{C,\rho} = & \oint_C\iint\frac{r_i}{r^2}\frac{\delta L_{C,\rho}}{\delta\sigma(\bm{x})}\frac{\delta L_{C,\rho}}{\delta\sigma(\bm{x}')}\mathrm{d}\sigma'\mathrm{d}x_i.
\end{align}
The first and second order area derivatives of $L_{C,\rho}$ are
\begin{equation}
\label{eq:dperi}
    \frac{\delta L_{C,\rho}}{\delta\sigma(\bm{x})} = 2\sqrt{2\pi}\rho\oint_C \epsilon_{ij}\partial_iG_\rho(\bm{x}-\bm{x}')\mathrm{d}x_j'
\end{equation}
and
\begin{equation}
    \frac{\delta^2 L_{C,\rho}}{\delta\sigma(\bm{x})\delta\sigma(\bm{x}')} = -2\sqrt{2\pi}\rho \partial_{kk}G_\rho(\bm{x}-\bm{x}'),
\end{equation}
respectively.
Thus, the first integration, $\mathcal{I}_{C,\rho}$, vanishes, since the integrand can be expressed as a gradient field
\begin{equation}
\begin{aligned}
    \frac{r_i}{r^2}\frac{\delta^2 L_{C,\rho}}{\delta\sigma(\bm{x})\delta\sigma(\bm{x}')} = \frac{2\sqrt{2\pi}}{\rho^4}\frac{\partial f}{\partial r_i}\left(-\frac{r^2}{2\rho^2}\right),
\end{aligned}
\end{equation}
where $f(x) = \operatorname{Ei}(x) + \exp(x)$ with $\operatorname{Ei}(x):=\int_{-x}^\infty t^{-1}{e^{-t}}\mathrm{d}t$ the exponential integral.

\begin{figure}[H]
    \centering
    \includegraphics{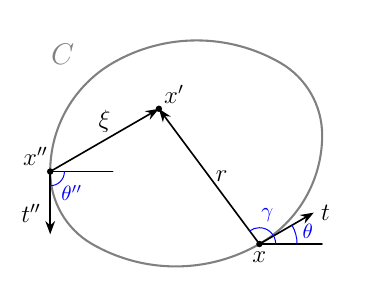}
    \caption{Illustrative plot of the angles $\theta,\theta'',\gamma$ and vectors $t_i,t_i'',r_i,\xi_i$.}
    \label{fig:illustrative plot}
\end{figure}

For the second integral, substituting (\ref{eq:dperi}) into $\mathcal{J}_{C,\rho}$, we obtain
\begin{equation}\label{J}
\begin{aligned}
   \mathcal{J}_{C,\rho} =& 2\sqrt{2\pi}\rho\oint_C\iint\oint_C\frac{\delta L_{C,\rho}}{\delta\sigma(\bm{x})}\frac{r_i}{r^2} \epsilon_{kl}\partial_kG_\rho(\bm{\xi})\mathrm{d}x_l''\mathrm{d}\sigma'\mathrm{d}x_i\\
   =&2\sqrt{2\pi}\rho\oint_C\oint_C\frac{\delta L_{C,\rho}}{\delta\sigma(\bm{x})}\underset{\zeta_{\rho}}{\underbrace{\iint\frac{r_it_it_l''}{r^2} \epsilon_{kl}\partial_kG_\rho(\bm{\xi})\mathrm{d}\sigma'}}\mathrm{d}s''\mathrm{d}s,
\end{aligned}
\end{equation}
where $r_i = x_i'-x_i$ and $\xi_i = x'_i-x''_i$.
Integrating $\zeta_{\rho}$ by parts, we have
\begin{equation}
    \begin{aligned}
    \zeta_{\rho} =& \underset{\text{boundary term (b.t.)}}{\underbrace{\iint \frac{\partial}{\partial x_k'}\left(\frac{\epsilon_{kl}t_it_l''r_i}{r^2}G_\rho(\bm{\xi})\right)\mathrm{d}\sigma'}}\\
    &- \iint G_{\rho}(\bm{\xi})\frac{\partial}{\partial r_k}\left(\frac{\epsilon_{kl}t_it_l''r_i}{r^2}\right)\mathrm{d}\sigma'\\
    =& \text{b.t.} + \iint\frac{\sin(\theta''+\theta-2\gamma)}{r^2}G_\rho(\bm{\xi})\mathrm{d}\sigma',
    \end{aligned}
\end{equation}
where $\theta,\theta'',\gamma$ are the angles of vectors $t_i,t_i'',r_i$ with respect to a fixed direction, say the $x$-axis, respectively (cf. Fig. \ref{fig:illustrative plot}).

The boundary term vanishes since $G_\rho$ decays exponentially as $x_i'$ goes to infinity.
For a fixed smooth loop, $\theta,\theta'',\gamma$ and $r$ are all smooth functions about $s,s''$ and $\xi_i$.
Considering
\begin{equation}
    \left.\frac{\partial \gamma}{\partial s''}\right|_{s,\xi_i} = \frac{\sin(\theta''-\gamma)}{r},\quad \left.\frac{\partial r}{\partial s''}\right|_{s,\xi_i} = \cos(\theta''-\gamma),
\end{equation}
we obtain
\begin{equation}
    \begin{aligned}
    \zeta_{\rho}  =& \iint\frac{\sin(\theta''+\theta-2\gamma)}{r^2}G_\rho(\bm{\xi})\mathrm{d}\sigma'\\
        =& \iint \left.\frac{\partial \psi}{\partial s''}\right|_{s,\xi_i} G_\rho(\bm{\xi})\mathrm{d}\sigma',
    \end{aligned}
\end{equation}
where
\begin{equation}
   \psi = -\frac{\sin(\theta-\gamma)}{r}.
\end{equation}
Thus, assuming that we can change the order of integration, integrating over $s''$, (\ref{J}) becomes
\begin{equation}
\mathcal{J}_{C,\rho} = 2\sqrt{2\pi}\rho\oint_C\iint \frac{\delta L_{C,\rho}}{\delta\sigma(\bm{x})}G_\rho(\bm{\xi}) \left(\oint_C \mathrm{d}\psi\right)\mathrm{d}\sigma'\mathrm{d}s = 0.
\end{equation}
Thus, the perimeter rule,
\begin{equation}
    P(C,\varGamma,t) = F(L_{C,\rho},\varGamma)
\end{equation}
is a steady state solution of the 2D loop equation.
\end{document}


\title{Supplementary Material for ``Area and Perimeter Rules of Velocity Circulation in Two-Dimensional Turbulence with Large-scale Absolute Equilibrium''}

\author{Zi-Ang Zhang}
\affiliation{College of Mechanics and Engineering Science and State Key Laboratory for Turbulence and Complex Systems, Peking University, Beijing 100871, China}

\author{Jin-Han Xie}
\email{jinhanxie@pku.edu.cn}
\affiliation{College of Mechanics and Engineering Science and State Key Laboratory for Turbulence and Complex Systems, Peking University, Beijing 100871, China}

\date{\today}

\begin{abstract}

\end{abstract}

\maketitle

\section{Area Rule for 2D Inviscid Loop Equation}

The regularized scalar area for a smooth non-self-intersecting loop $C$ in 2D plane could be defined as,
\begin{equation}
   A_{C,\rho}=\iint_{S_C}\iint_{S_C}G_\rho(\bm{x}-\bm{x}')\mathrm{d}\sigma(\bm{x}')\mathrm{d}\sigma(\bm{x}),
\end{equation}
where $G_\rho(\bm{r})$ is the 2D Gaussian kernel,
\begin{equation}
    G_\rho(\bm{r}) = \frac{1}{2\pi\rho^2}e^{-\frac{r^2}{2\rho^2}}.
\end{equation}
For arbitrary finite radius of regularization $\rho$, $A_{C,\rho}$ is an smooth quadratic functional about loop $C$, thus, its area derivative is well-defined, which reads,
\begin{equation}
    \frac{\delta A_{C,\rho}}{\delta\sigma(\bm{x})} = 2\iint_{S_C}G_\rho(\bm{x}-\bm{x}')\mathrm{d}\sigma(\bm{x}'),
\end{equation}
and the second order derivative reads,
\begin{equation}
    \frac{\delta^2 A_{C,\rho}}{\delta \sigma(\bm{x})\delta\sigma(\bm{x}')} = 2G_\rho(\bm{x}-\bm{x}').
\end{equation}
Specially, for $\bm{x}\in C$, applying the local expansion of smooth loop $C$, we have,
\begin{equation}
    \frac{\delta A_{C,\rho}}{\delta\sigma(\bm{x})} = 1 + \kappa\rho + O(\rho^2),
\end{equation}
where $\kappa$ is the curvature at point $\bm{x}\in C$.
We show in this section that the circulation PDF,
\begin{equation}
\label{eq:area rule}
    P(C,\varGamma,t) = F(A_{C,\rho},\varGamma),
\end{equation}
is a steady solution to the 2D inviscid loop equation,
\begin{equation}
    \label{eq:Two dimensional loop equation}
    \frac{\partial^2P(C,\varGamma,t)}{\partial \varGamma\partial t} = \oint_C\iint \frac{r_i}{2\pi r^2}\frac{\delta^2 P(C,\varGamma,t)}{\delta\sigma(\bm{x})\delta\sigma(\bm{x}')}\mathrm{d}\sigma(\bm{x}')\mathrm{d}x_i,
\end{equation}
for $F$ being an arbitrary smooth function when $\rho\to 0$.
Here, $r_i = x'_i-x_i$
Substituting (\ref{eq:area rule}) into (\ref{eq:Two dimensional loop equation}), we have,
\begin{widetext}
\begin{equation}
\label{eq:area loop equation}
\begin{aligned}
    \frac{\partial^2 P}{\partial \varGamma\partial t} =& \frac{\partial F}{\partial A_{C,\rho}}\underset{\mathcal{I}_{C,\rho}}{\underbrace{\oint_C\left(\iint\frac{\bm{r}}{2\pi r^2}\frac{\delta^2 A_{C,\rho}}{\delta\sigma(\bm{x})\delta\sigma(\bm{x}')}\mathrm{d}\sigma(\bm{x}')\right)\cdot\mathrm{d}\bm{x}}}+\frac{\partial^2 F}{\partial A_{C,\rho}^2}\underset{\mathcal{J}_{C,\rho}}{\underbrace{\oint_C\left(\iint\frac{\bm{r}}{2\pi r^2}\frac{\delta A_{C,\rho}}{\delta\sigma(\bm{x})}\frac{\delta A_{C,\rho}}{\delta\sigma(\bm{x}')}\mathrm{d}\sigma(\bm{x}')\right)\cdot\mathrm{d}\bm{x}}},
\end{aligned}
\end{equation}
\end{widetext}
where, the integration $\mathcal{I}_{C,\rho}$ vanishes, since,
\begin{equation}
    2\frac{\bm{r}}{r^2}G_\rho(\bm{r}) = \frac{1}{2\pi\rho^2}\nabla\operatorname{Ei}\left(-\frac{r^2}{2\rho^2}\right),
\end{equation}
where $\operatorname{Ei}$ is the exponential integration,
thus its integration over a closed loop $C$ vanishes.

Substitute the area derivative of $A_{C,\rho}$ in $\mathcal{J}_{C,\rho}$,
we have,
\begin{equation}
\begin{aligned}
    \mathcal{J}_{C,\rho} =& \oint_C\iiiint \frac{r_i}{2\pi r^2}(1+\kappa\rho + O(\rho^2)) G_\rho(\bm{r}')\mathrm{d}\sigma''\mathrm{d}\sigma'\mathrm{d}x_i,
\end{aligned}
\end{equation}
where $r_i = x'_i-x_i$, $r'_i = x''_i-x'_i$ and $\mathrm{d}\sigma'$, $\mathrm{d}\sigma''$ are the infinitesimal area element at points $\bm{x}',\bm{x}''$, respectively. 
Exchange the order of integration, we first integrate over the closed loop $C$, which gives,
\begin{equation}
    \oint_C \frac{r_i}{2\pi r^2} (1+\kappa\rho + O(\rho^2))\mathrm{d}x_i = \rho\oint\frac{\kappa r_i}{2\pi r^2}\mathrm{d}x_i + O(\rho^2),
\end{equation}
which converges to zero when $\rho\to 0$.
Thus, the area rule,
\begin{equation}
    P(C,\varGamma,t) = \lim_{\rho\to 0}F(A_{C,\rho},\varGamma)
\end{equation}
is a steady solution to the 2D loop equation (\ref{eq:Two dimensional loop equation}).

\newpage
\begin{widetext} 
\section{Suplementrary Data}
In FIG.\ref{fig:placeholder}, we plotted the integrand of the 10th order moments of the velocity circulation to verify that the amount of samples is adequate.
\begin{figure*}[htbp]
    \centering
    \includegraphics{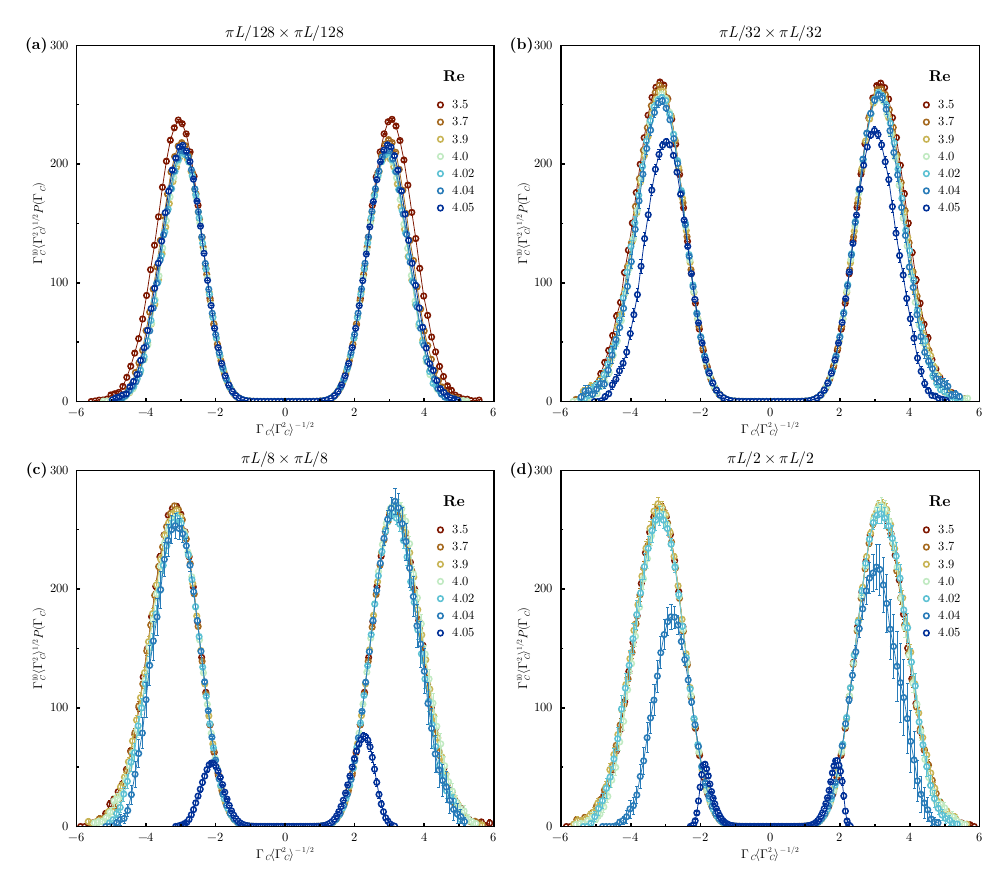}
    \caption{The integrand of the 10th-order moments of velocity circulation over square loops with side lengths $\pi/128,\pi/32,\pi/8$ and $\pi/2$. The Reynolds number of each curve is encoded by its color illustrated the top-left panel.}
    \label{fig:placeholder}
\end{figure*}

\end{widetext}
